\documentclass[12pt]{iopart}
\usepackage{times}
\usepackage{graphicx}

\begin{document}

\title{Quantum computing using dissipation}
\author{A Beige}
\address{Blackett Laboratory, Imperial College London, Prince Consort Road, London, SW7 2BW, UK}
\begin{abstract}
The principal obstacle to quantum information processing with many qubits is decoherence. One source of decoherence is spontaneous emission which causes loss of energy and information. Inability to control system parameters with high precision is another possible source of error. As a solution we propose quantum computing experiments {\em using} dissipation based on an environment-induced quantum Zeno effect. As an example we present a simple scheme for quantum gate implementations with cold trapped ions in the presence of cooling.
 \end{abstract}

\section{Introduction}

Following the theoretical formulation of quantum computing \cite{Deutsch86} and the first algorithms for problems which can be solved more easily on a quantum computer than on a classical computer \cite{Shor:94,Grover97}
the practical implementation of such a device has become a challenging task. Initial steps have already been taken. Quantum bits (qubits) can be realised for instance by storing the information in a superposition of the internal 
states of two-level atoms. However building systems with many coupled qubits remains a huge challenge.  Many demands must be met: reliable qubit storage, preparation and measurement, gate operations with high fidelity and low failure rate and scalability of the system to many qubits.  The biggest problem is posed by decoherence which can cause the loss of information.

In this paper we review the idea of quantum computing using dissipation \cite{Letter,Ben} which might help to overcome the decoherence problem in some systems. On the contrary, it might even be useful to introduce an additional spontaneous decay channel into a system in order to create a realm of new possibilities to implement gate operations between qubits. To illustrate this we present a scheme for the realisation of quantum computing between cold trapped ions in the presence of cooling. The ions are stored inside a linear trap and each qubit is obtained from two different ground states of the same ion, as in \cite{simple,cold}. 

In the quantum computing scheme proposed here, the role of cooling during gate operations is twofold. On one hand it decreases the sensitivity of the scheme with respect to heating \cite{eschner}. On the other hand, the presence of the cooling lasers introduces a decay channel into the system whose presence leads to a restriction of the time evolution of the system onto the computational subspace and facilitates the possibility for quantum gate implementations within one step. The probability for photon emission is relatively small since the system remains to a good approximation within a decoherence-free subspace \cite{Palma,Zanardi97,Beigenjp}. If heating nevertheless populates the vibrational mode or gate failure moves the system out of the decoherence-free subspace, then photons are emitted at a high rate  \cite{wine79,eschner,leib}. This can be de\-tec\-ted and the computation can be restarted. 

Many schemes for quantum computing with trapped ions have already been proposed. Some of them require cooling of the ions into the ground state of a common vibrational mode. That this is possible has been demonstrated recently in Innsbruck \cite{nature1}, where the Cirac-Zoller controlled-NOT quantum gate  \cite{cz} has been implemented with the help of six concatenated laser pulses individually addressing each of the two ions.  At the same time, the group in Boulder demonstrated a robust, high-fidelity geometric two ion-qubit phase gate in the laboratory. This was achieved with a sequence of laser pulses and without individual addressing of the ions  \cite{nature2}.  But are these schemes really suitable for quantum computation with many qubits? Finding reliable ways to scale present schemes to many qubits might require further simplifications of the experimental setup without decreasing the precision of gate operations. 

\section{Non-Hermitian Hamiltonians and no-photon time evolutions}

Before we discuss the coherent control of an open quantum system with the help of dissipation in more detail, let us first introduce the theoretical model for describing the time evolution of the system under the condition of no photon emission. We give a short review of the quantum jump method \cite{HeWi11} which is equivalent to the Monte Carlo wave-function \cite{HeWi2} and the quantum trajectory \cite{HeWi3} approaches \cite{review}.

\subsection{The quantum jump approach}

In the last decades, several quantum optics experiments have been performed studying the statistics of photons emitted by {\em single} quantum mechanical systems, like one or two laser-driven trapped atoms or ions. Effects have been found that would be averaged out in the statistics of photons emitted by a whole {\em ensemble} and which cannot be predicted with the help of expectation values calculated for a statistical ensemble \cite{Hegerfeldt}. New formalisms describing single systems interacting with the environment had to be developed.

A typical example for such an experiment is electron shelving \cite{shelving}, i.e. the occurance of stochastic macroscopic light and dark periods in the resonance fluorescence of a laser-driven atom with a metastable state. Another example is the two-atom double-slit experiment by Eichmann {\em et al.} \cite{Eichmann} which demonstrated that the photons emitted by two atoms at a fixed distance can create an interference pattern on a distant screen. These experiments suggest that the effect of the environment on the state of the atoms is the same as the effect of rapidly repeated measurements of whether a photon has been emitted or not \cite{behe,schoen}. From this assumption the quantum jump approach has been derived. 

Assume that a measurement is performed on a quantum optical system surrounded by a free radiation field initially in its vaccum state $|0_{\rm ph} \rangle$ and prepared in $|\psi \rangle$ determining after a time $\Delta t$ whether or not a photon has been created.  If $H$ is the Hamiltonian including the interaction of the system with its environment, the state of the system equals
\begin{equation} \label{def}
|0_{\rm ph} \rangle \, U_{\rm cond}(\Delta t,0) |\psi \rangle
\equiv    |0_{\rm ph} \rangle  \langle 0_{\rm ph} | \, U(\Delta t,0) \, |0_{\rm ph} \rangle |\psi \rangle 
\end{equation}
under the condition that the free radiation field is still in the vacuum state. For quantum optical experiments, it has been shown that the dynamics under the conditional time evolution operator $U_{\rm cond}(\Delta t,0)$, defined by the right hand side of (\ref{def}), can be summarised by a Hamiltonian $H_{\rm cond}$ that is largely independent of the choice of $\Delta t$. The conditional Hamiltonian $H_{\rm cond}$ is non-Hermitian and the norm of a state vector developing with $H_{\rm cond}$ decreases in general in time. Under the condition of no photon emission, the state of the system equals at time $t$
\begin{equation} \label{seq} 
|\psi^0(t)\rangle = U_{\rm cond}(t,0) \, |\psi \rangle/\| \cdot \| ~.
\end{equation}
For convenience, $H_{\rm cond}$ has been defined such that
\begin{equation} \label{P0}
P_0(t,\psi) = \| \, U_{\rm cond} (t,0) \, |\psi \rangle \, \|^2 
\end{equation}
is the probability for no photon emission in $(0,t)$. 

The non-Hermiticity of the conditional Hamiltonian $H_{\rm cond}$ and the continuous decrease of the amplitude of the unnormalised state vector $U_{\rm cond} (t,0) \, |\psi \rangle$ reflect that the observation of no photons reveals information about the system. The longer no photon is emitted the more unlikely it becomes that there is excitation that might cause an emission and the amplitudes of states with spontaneous decay rates decrease exponentially  \cite{cook}.

\subsection{Example: Two four-level ions inside a linear ion trap} \label{trap}

\begin{figure}
\begin{center}
\includegraphics{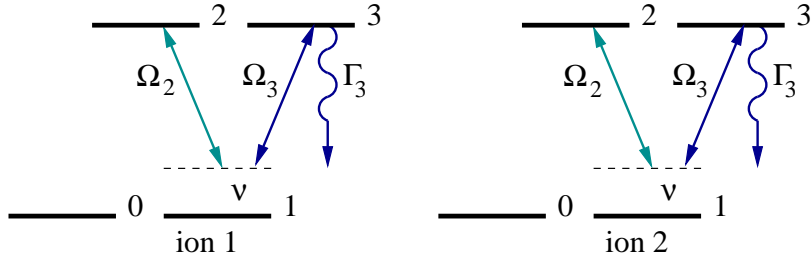}
\end{center}
\caption{Level scheme of the two ions involved in the gate operation. Each qubit is obtained from the ground states $|0\rangle$ and $|1\rangle$ of one ion. In addition a metastable state $|2\rangle$, a rapidly decaying state $|3\rangle$ with decay rate $\Gamma_3$ and two strong laser fields with coupling strength $g_j= {1 \over 2} \, \eta_j \Omega_j$ and detuning $\nu$ are required.} \label{fig1}
\end{figure}

The level scheme of the two cold trapped ions that we consider in this paper in order to discuss the role of dissipation in quantum gate realisations is shown in Figure \ref{fig1}. Each qubit is obtained from two different atomic ground states $|0 \rangle$ and $|1 \rangle$ of the same ion. In addition, a metastable state $|2 \rangle$ and a rapidly decaying level 3 are required. First, the ions have to be cooled into the ground state of a common vibrational mode.  Two strong laser fields detuned by the frequency $\nu$ of a common vibrational mode should be applied. The laser field coupling to the 1-2 transition  establishes a coupling between the two qubits involved in the gate operation. The laser field driving the 1-3 transition represents the laser cooling setup and can be replaced by any other laser cooling configuration without changing the effective time evolution of the system.

In the following, we denote the spontaneous decay rate of level 3 by $\Gamma_3$ while $b$ and $b^\dagger$ are the annihilation and creation operator of a phonon in the common vibrational mode. The coupling constant of this mode to the atomic 1-$j$ transition equals $g_j \equiv {1 \over 2} \, \eta_j \Omega_j$ where $\Omega_j$ is the Rabi frequency of the applied laser field and $\eta_j$ is the Lamb Dicke parameter depending on the characteristics of the ion trap. Proceeding as in  \cite{HeWi11}, one finds that the conditional Hamiltonian within the dipole and the rotating wave approximation and in the interaction picture with respect to the free Hamiltonian equals
\begin{eqnarray} \label{hlaserI}
H_{\rm cond} &=& \sum_{i=1}^2 {\rm i} \hbar \, \big[ \, g_2 \, |1 \rangle_i\langle 2| \, b^\dagger 
+ g_3 \, |1 \rangle_i\langle 3| \, b^\dagger - {\rm h.c.} \, \big]
- \sum_{i=1}^2 {\textstyle{{\rm i} \over 2}} \hbar \Gamma_3 \, |3 \rangle_i\langle 3| ~.
\end{eqnarray}
Here the Lamb-Dicke regime and the condition $\nu \gg \Omega_2$,  $\Omega_3$ has been assumed, as in  \cite{cz}. 

\section{Quantum computing using dissipation}

The basic idea of {\em quantum computing using dissipation} is to utilise the {\em coherent} no-photon time evolution given by $H_{\rm cond}$ for the implementation of gate operations. Whenever a photon emission occurs the computation fails and has to be repeated. Nevertheless, we show that fidelities $F$ and success rates $P_0$ close to one can be achieved for a wide range of experimental parameters. The reason that spontaneous emission from the ions is in the following negligible is that the system remains during the whole computation in a decoherence-free (DF) state \cite{Palma,Zanardi97,Beigenjp}. In this section we define DF states and describe a mechanism which restricts the time evolution of a system onto its decoherence-free subspace (DFS).

\subsection{Decoherence-free states} \label{DFSSS}

The DFS of a system is a subspace of states whose population does not lead to decoherence. Using the quantum jump approach, a state $|\psi \rangle$ is DF if $P_0(t,\psi) = 1$ for all times $t$ \cite{Beigenjp}. Hence, the DFS is spanned by the eigenvectors of the conditional Hamiltonian with real eigenvalues $\lambda_i$. The eigenvectors $|\lambda_i \rangle$ of $H_{\rm cond}$ are in general non-orthogonal. It is therefore useful to introduce the reciprocal basis vectors $|\lambda^j \rangle$ with $\langle \lambda^j |\lambda_i \rangle = \delta_{ij}$ and to write the conditional Hamiltonian as 
\begin{eqnarray}
H_{\rm cond} &=& \sum_i \lambda_i \, |\lambda_i \rangle \langle \lambda^i|~.
\end{eqnarray}
Suppose that all non-DF states couple strongly to the environment and populating them leads typically to a photon emission within a time $\Delta t$. The time $\Delta t$ is greater than a certain minimal size which can be determined from the quantum jump approach. Provided that the eigenvalues $\lambda_k$ corresponding to 
non-DF states fulfil the condition
 \begin{equation} \label{ass}
{\rm e}^{-{\rm i} \lambda_k \Delta t/\hbar} = 0 ~,
\end{equation}  
the no-photon time evolution operator becomes  \cite{remark} 
\begin{equation} \label{ups}
U_{\rm cond}(\Delta t,0) = \sum_{i: |\lambda_i \rangle \in {\rm DFS}} 
{\rm e}^{-{\rm i} \lambda_i \Delta t/\hbar} \, |\lambda_i \rangle \langle \lambda_i|  ~.
\end{equation}
This operator projects every initial state onto the DFS. The action of the environment over a time $\Delta t$ can therefore be interpreted as a measurement whether the system is DF or not. The probability for no emission in $\Delta t$ equals the probability to be in a DF state. 

\subsection{An environment-induced quantum Zeno effect}

These continuous measurements caused by the environment lead to a realm of possibilities to manipulate a system within the DFS. Actually, any arbitrary interaction can be used as long as its typical time scale is much longer than $\Delta t$ defined by condition (\ref{ass})  \cite{Letter}. This restriction of the system onto the DFS can intuitively be understood with the help of the quantum Zeno effect  \cite{misra}. Within $\Delta t$, a weak interaction can only transfer population proportional $\Delta t$ out off the DFS. During the next measurement the system is found in a non-DF state with a probability proportional $\Delta t^2$. Otherwise it is projected back onto the DFS by the time evolution operator (\ref{ups}). Important is that the probability to find the system always in a DF state, i.e. $T/\Delta t$ times if $T$ is the gate operation time, goes in the limit of weak interactions, i.e. for $\Delta t/T \to 0$, to one. 

However, the time evolution of the system inside the DFS is not inhibited. Using first order perturbation theory with respect to the weak interaction, (\ref{ups}) and the assumption that the system is initially in a DF state, one can show that the conditional time evolution in $\Delta t$ is the same as the one that one obtains from the Hamiltonian
\begin{equation} \label{heff}
H_{\rm eff} = I\!\!P_{\rm DFS} \, H_{\rm cond} \, I\!\!P_{\rm DFS}
\end{equation}
using the same approximations. Here $I\!\!P_{\rm DFS} = \sum_{i:|\lambda_i\rangle \in {\rm DFS}} |\lambda_i\rangle \langle \lambda_i|$ is the projector onto the DFS.  The effective Hamiltonian $H_{\rm eff}$ is used in the following to find the appropriate laser configuration for the realisation of certain gate operations.

One motivation among others to use an environment-induced quantum Zeno effect for quantum computing is the simplicity of the resulting schemes. As we see in the next section, the DFS of the cold-ion system contains in addition to all ground states, highly entangled states. Through populating these states entanglement between qubits can be created during the effective time evolution even if this is in general not possible using only a single laser pulse. 
 
\subsection{Scaling of probabilistic quantum computing schemes}
 
Let us now estimate the probability of finding the result of a whole computation assuming that each gate can be performed with maximum fidelity (as it applies as long as the corrections to the desired effective time evolution are not too big \cite{cold}) but only with a finite success rate $P_0$. The probability of implementing an algorithm of $N$ gates faultlessly is $P_0^N$ and grows exponentially with $N$. On the other hand, if one always knows whether an algorithm has failed or not, the computation can be repeated until a result is obtained. The probability for not having a result after $M$ runs equals
\begin{equation}
P_{\rm no \, result}= \big(\, 1 - P_0^N \, \big)^M ~.
\end{equation}
For large $N$ this is approximately $\exp \big(- M \,  P_0^N \big)$. Many repetitions might be necessary to implement a computation. However, for smaller numbers of $N$ and if $P_0$ is sufficiently close to unity, the failure probability is already nearly negligible for $M \approx N$. For example, if $P_0=95\,\%$ and an algorithm with $N=50$ gates is performed, then repeating the computation 50 times yields a success rate above $98\,\%$. 

\section{Quantum computing with cold trapped ions in the presence of cooling} \label{rough}

We now consider again the concrete setup described in subsection \ref{trap} and describe the realisation of two-qubit quantum gates between cold trapped ions in the presence of cooling with the help of additional weak laser fields.

\subsection{Decoherence-free states of the ions}

To predict the effect of these fields we first determine the decoherence-free (DF) states of the system in the absence of any additional interaction. As shown in subsection \ref{DFSSS}, they can be calculated by finding the eigenvectors of the Hamiltonian (\ref{hlaserI}) with real eigenvalues. This yields that the decoherence-free subspace (DFS) of the system contains only superpositions of states with the ions either in the state $|00\rangle$ combined with an arbitrary state of the vibrational mode or both ions in $|01\rangle$, $|10\rangle$, $|11\rangle$ or in the antisymmetric and maximally entangled state
\begin{equation} \label{a}
|a\rangle \equiv {\textstyle {1\over \sqrt{2}}} \, \big[ |12 \rangle -|21\rangle \big]
\end{equation}
while the vibrational mode is not populated. Here all real eigenvalues are zero and the system does not evolve as long as no additional interaction is applied. 

\subsection{Single laser pulse quantum gates}

To realise gate operations between the two qubits formed by the ground states of the ions weak laser fields are applied in addition to the strong lasers shown in Figure \ref{fig1}. Let us denote the Rabi frequency of the laser with respect to the $j$-2 transition in ion $i$ by $\Omega_j^{(i)}$ and assume that                                                                          
\begin{equation}
\Omega_j^{(i)} \ll g_2, ~ g_3 ~~ {\rm and} ~~ \Gamma_3~.
\end{equation}
The conditional Hamiltonian $H_{\rm cond}$ of the system then becomes the sum of the Hamiltonian (\ref{hlaserI}) and the laser Hamiltonian 
\begin{eqnarray} \label{hint}
H_{\rm laser} &=& \sum_{i=1,2} \sum_{j=0,1}  {\textstyle {1 \over 2}} \hbar \Omega_j^{(i)} \, |j \rangle_i\langle 2| + {\rm h.c.} 
\end{eqnarray}
If the time evolution of the system is restricted onto the DFS, as predicted in the previous section, then it is effectively given by the Hamiltonian (\ref{heff}) which equals
\begin{eqnarray} \label{heff2}
H_{\rm eff} &=& {\textstyle {1 \over 2 \sqrt{2}}} \hbar \, \big[ \, - \Omega_0^{(1)} \, 
|01 \rangle + \Omega_0^{(2)} \, |10 \rangle 
+ \big( \Omega_1^{(2)} - \Omega_1^{(1)} \big) \, |11 \rangle \, \big] \, \langle a| + {\rm h.c.} 
\end{eqnarray}
This Hamiltonian can be used to implement quantum gate operations if the operation time $T$ is chosen such that at the end of each gate again only qubit states are populated. A more detailed analysis of the no-photon time evolution of the ions is given in \cite{cold}.

\subsection{The controlled-NOT gate}

As a concrete example we consider the controlled NOT (CNOT) gate. If ion 1 contains the target qubit and ion 2 provides the control qubit this gate corresponds to the time evolution operator
\begin{equation}\label{ucnot}
U_{\rm gate} = |00\rangle \langle00| + |01\rangle \langle01| + |10\rangle \langle11| + |11\rangle \langle10| ~.
\end{equation}
The easiest way to realise a CNOT gate is to couple one laser with the (real) Rabi frequency $\Omega \equiv \Omega_1^{(1)}$ to the 1-2 transition of ion 1 and another one with the same Rabi frequency to the 0-2 transition of ion 2 which yields the effective Hamiltonian
\begin{eqnarray} \label{ex1}
H_{\rm eff} &=& {\textstyle {1 \over 2 \sqrt{2}}} \hbar \Omega \, \big[ \, |10 \rangle - |11 \rangle \, \big] \, \langle a|  + {\rm h.c.} 
\end{eqnarray}
If the duration of the laser pulse equals $T = 2\pi/\Omega$ the resulting time evolution can be shown to be exactly the desired operation \cite{cold}.

\begin{figure}
\begin{minipage}{\columnwidth} 
\begin{center}
\resizebox{\columnwidth}{!}{\rotatebox{0}{\includegraphics{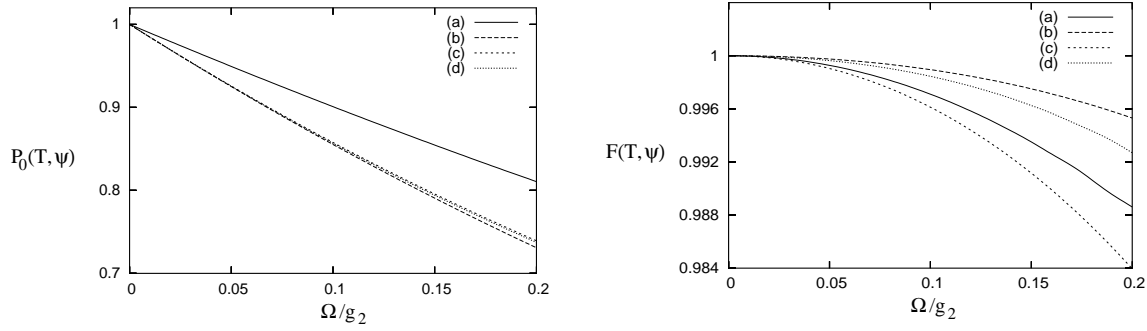}}}
\end{center}
\caption{Success rate and fidelity under the condition of no photon emission of a single CNOT gate as a function of the Rabi frequency $\Omega$ for $\Gamma_3 = 2 \sqrt{37} \, g_2$, $g_3 = \sqrt{2} \, g_2$  and for the initial qubit states $|00 \rangle$ (a), $|10 \rangle$  (b),  $|11 \rangle$ (c) and $[ \, |10\rangle - |11 \rangle \, ]/\sqrt{2}$ (d). The gate success rate and fidelity are always maximal if the ions are initially prepared in $|01 \rangle$.}\label{fig3}
\end{minipage}
\end{figure}

Figure \ref{fig3} results from a numerical solution of the no-photon time evolution given by the sum of the Hamiltonians (\ref{hlaserI}) and (\ref{hint}). For very small Rabi frequencies, the gate success rate and fidelity is for all initial states close to one. For larger values of $\Omega$, the $P_0(T,\psi)$ decreases and is for $\Omega = 0.2\, g_2$ as low as $73 \,\%$. The gate fidelity is in this case still above $98.4\,\%$. The smallest gate success rate is found when the atoms are initially in $|10 \rangle$, $|11 \rangle$ or in a superposition of these two states. In general, success rates $P_0 > 90\,\%$ are achieved as long as $\Omega < 0.07 \,g_2$. 

\begin{figure}
\begin{center}
\includegraphics{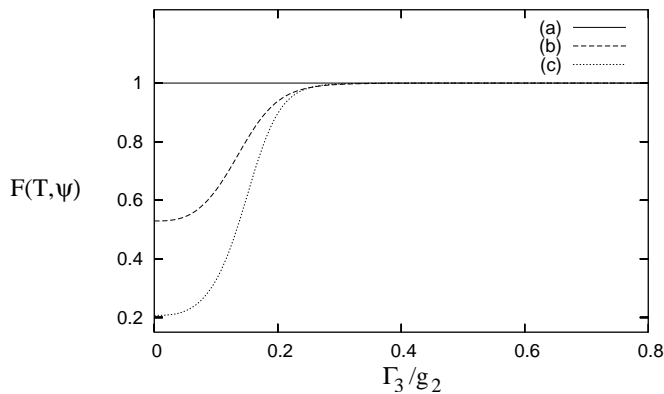}
\end{center}
\vspace*{-0.5cm}
\caption{Fidelity for a single CNOT gate under the condition of no photon emission as a function of the spontaneous decay rate $\Gamma_3/g_2$ for $g_3=\Gamma_3$, $\Omega=0.01 \, g_2$ and for the initial qubit states $|00 \rangle$ (a), $|10 \rangle$  (b) and $[ \, |10\rangle - |11 \rangle \, ]/\sqrt{2}$ (c). If the ions are initially prepared in $|01 \rangle$, the gate success rate equals one. For $|\psi \rangle = |10 \rangle$ the fidelity is about the same as in graph (c).}\label{fig22}
\end{figure}

Finally, we would like to comment on the role of the spontaneous decay rate of level 3 in the scheme. Figure \ref{fig22} shows the fidelity of a single CNOT gate as a function of $\Gamma_3$ and for $g_3=\Gamma_3$. In the chosen parameter regime, the effective damping rate of unwanted population in non-DF states can be shown to equal $g_3^2/\Gamma_3 = \Gamma_3$ to a very good approximation and the effective decay rate of non-DF states increases linearly in $\Gamma_3$. Figure \ref{fig22} confirms that the presence of the auxiliary dissipation channels is crucial for the scheme to work! For small damping rates the mechanism which restricts the time evolution of the system onto the computational subspace fails and the minimum gate fidelity is well below $50 \, \%$. 

\section{Conclusions}

The paper reviews the idea of quantum computing using dissipation which offers a great variety of possibilities to implement gate operations in the presence of only one dissipation channel in a system. As an example we described the possibility to implement precise quantum gates between cold trapped ions in the presence of cooling of a common vibrational mode and within one step. Because of its simplicity and robustness against parameter fluctuations \cite{cold}, the proposed scheme might help to increase the number of qubits in present quantum computing experiments. \\
  
{\em Acknowledgement.} This work was supported by the Royal Society in form of a University Research Fellowship and by the EPSRC and the European Union in part.


\begin{thebibliography}{9}
\bibitem{remark}
The eigenvectors $|\lambda_i \rangle$ of $H_{\rm cond}$ which correspond to DF states are orthogonal to each other and one has $|\lambda^i \rangle = |\lambda_i \rangle$.
\end{thebibliography}
\end{document}